\documentclass[twocolumn,12pt]{article}

\usepackage[colorlinks=true,linkcolor=blue]{hyperref}
\usepackage{amsmath}
\usepackage{graphicx}
\usepackage{subcaption}
\usepackage{stackengine}
\usepackage{indentfirst}
\usepackage{ulem}
\usepackage{cite}
\usepackage{float}
\usepackage{chemformula}
\usepackage[affil-it]{authblk}

\newcommand{\bq}{\mathbf{q}} 
\newcommand{\ANG}{\mathring{\mathrm{A}}}
\newcommand{\BF}{\mathbf{F}} 
\newcommand{\BG}{\mathbf{G}} 
\newcommand{\BA}{\mathbf{A}} 

\title{Three Dimensional Imaging using Coherent X Rays at Grazing Incidence Geometry}
\author[1]{Yi Yang}
\author[1,*]{Sunil K. Sinha}
\affil[1]{Department of Physics, University of California—San Diego, La Jolla, California 92093, USA}
\affil[*]{\url{ssinha@physics.ucsd.edu}}
\date{}

\begin{document}

\twocolumn[
\begin{@twocolumnfalse}
\maketitle
\begin{abstract}
We have developed a 3 dimensional Coherent Diffraction Imaging (CDI) algorithm to retrieve phases of diffraction patterns of samples in Grazing Incidence Small Angle X-ray Scattering (GISAXS) experiments. The algorithm interprets the diffraction patterns using the Distorted-Wave Born Approximation (DWBA) instead of the Born Approximation (BA), as in this case the existence of a reflected beam from the substrate causes the diffraction pattern to deviate significantly from the simple Fourier transform of the object. Detailed computer simulations show that the algorithm works. Verification with real experiments is planned.
\end{abstract}
\end{@twocolumnfalse}
]
\section{Introduction}
Coherent Diffraction Imaging (CDI) with X-Rays has become an important area of synchrotron-based research, with promising applications in materials science, nanoscience, and the biological sciences.  The technique can be applied to image objects down to the 10nm length scale. It is based on retrieving the phase information from an oversampled~\cite{gerchberg1972practical,miao1998phase} diffraction pattern using iterative methods. Most CDI studies have been carried out in transmission geometry in 2 dimensions (2D)~\cite{miao1999extending}. In such a case, the 2D diffraction pattern is the Fourier transform of the sample normal to the beam direction, according to the Born Approximation (BA).

Three dimensional (3D) CDI in transmission geometry is also possible via tomography by rotating the sample multiple times about an axis normal to the beam during scattering. The first experimental demonstration of 3D CDI was published by Miao et.al~\cite{miao2002high}, in which a 3D diffraction pattern was assembled from a series of 2D diffraction patterns. Later on, Chapman et.al~\cite{chapman2006high}  used a similar method to reconstruct 3D images \textit{ab initio}. In the same year, Miao et.al ~\cite{miao2006three} succeeded in getting multiple 2D projected images in object space using 2D CDI at different rotation angles and tomographically computing them to a 3D image. For single crystals, 3D CDI was extended to map out the internal strain field inside the crystal~\cite{pfeifer2006three}. For a review of CDI, see Ref.~\cite{miao2011coherent}.

However, there are several cases where it is desirable to employ GISAXS, such as where the sample is deposited on an opaque substrate, as in the case of a thin film, or on a liquid surface and to scatter in reflection geometry at low angles of incidence into a 2D detector (Fig.~\ref{fig:samplesetup}).
The GISAXS method has become increasingly popular for studying systems such as quantum dots and nanoparticles in thin films, supported catalyst materials, integrated circuits, etc. (For a review, see Ref.~\cite{renaud2009probing}.)

A problem arises for scattering at angles close to or smaller than the critical angle for total reflection of the substrate, which is how to deal with the problem of the specularly reflected beam from the substrate which interferes coherently with the scattering from the desired object. 
The strong specularly reflected beam from the substrate produces double scattering effects which modify the off-specular scattering from the sample. A reasonably successful approximate way of taking these effects into account has been to use the Distorted-Wave Born Approximation (DWBA)~\cite{sinha1988x,renaud2009probing,jiang2011waveguide}.
 In this paper, we address the problem of retrieving the phases of the diffraction patterns using the DWBA rather than the simple BA and applying it to the CDI algorithms.

\section{Distorted-Wave Born Approximation}
Given the electron density $f(\mathbf{r})$ of an object, its Fourier transform $F(\mathbf{q})$ is:

\begin{equation}
\label{eqn:ft}
F(\mathbf{q})=\sum_{x=0}^{N_x-1}\sum_{y=0}^{N_y-1}\sum_{z=0}^{N_z-1}f(\mathbf{r})
e^{-2\pi i(\frac{q_x x}{N_x}+\frac{q_y y}{N_y}+\frac{q_z z}{N_z})}
\end{equation}

\noindent where $\mathbf{r}=(x,y,z)$, the spatial coordinates in object space, and $\mathbf{q}=(q_x,q_y,q_z)$, the wavevector transfer in reciprocal space. $\bq$ and $\mathbf{r}$ are discretized and range in each dimension from 0 to $N_x-1,N_y-1,N_z-1$, respectively. $f(\mathbf{r})$ and $F(\bq)$ are periodic with a period of $N_x,N_y,N_z$ in the $x,y,z$ and $q_x,q_y,q_z$ directions, respectively. In the context, the range in the $z$ and $q_z$ direction may also be written as $[-\frac{N_z}{2},\frac{N_z}{2}-1]$, which is equivalent to $[0,N_z-1]$.

The Born Approximation (BA) states that the differential cross section ($\frac{d\sigma}{d\Omega}$) is:

\begin{equation}
\label{eqn:ba}
\frac{d\sigma}{d\Omega}=r_e^2|F(\mathbf{q})|^2
\end{equation}
\noindent where $r_e$ is the Thompson scattering length and $F(\bq)$ is also known as the BA form factor.

\begin{figure}[htbp]
\centering

\begin{subfigure}{.5\textwidth}
\includegraphics[scale=.4]{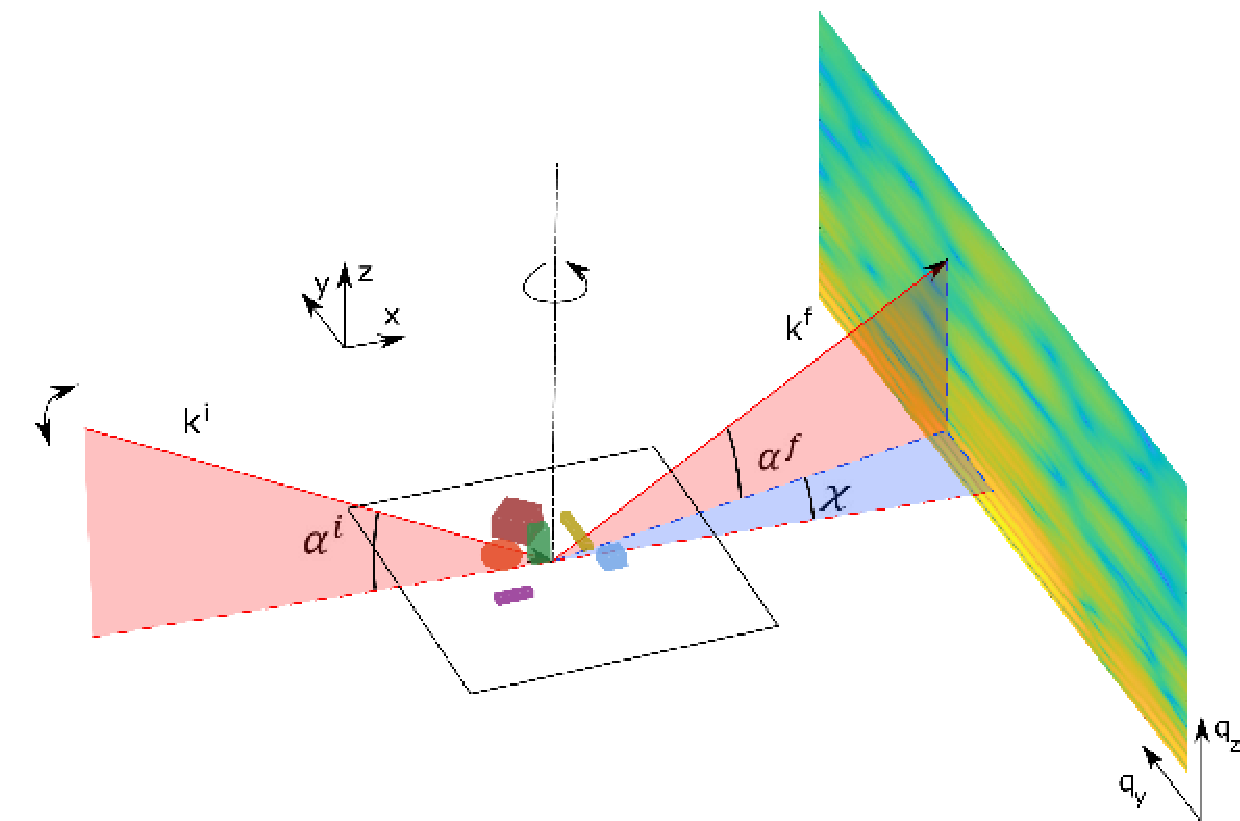}
\end{subfigure}

\caption{ Grazing Incidence Small-Angle Scattering layout. $\protect \mathbf{k}^i$ and $\protect \mathbf{k}^f$ are the incident and scattered wavevectors. The angles $\protect \alpha^i$, $\protect \alpha^f$ and $\protect \chi$ are related to the wavevector transfer. 
}
\label{fig:samplesetup}
\end{figure}

In reflection geometry, at angles close to or smaller than the critical angle of the substrate, the BA breaks down, as the specularly reflected beam from the substrate interferes coherently with the scattering from the desired object. We assume that the objects to be imaged are contained within a single layer on top of a smooth substrate and introduce the single-layer DWBA diffuse (specular part is omitted because we assume that it is blocked by the beamstop) differential cross section expression~\cite{jiang2011waveguide}

\begin{equation}
(\frac{d\sigma}{d\Omega})_{diff}\approx r_e^2|G(\mathbf{q}_{\parallel},\alpha^i,\alpha^f)|^2
\end{equation}
\noindent where $\alpha^i$, $\alpha^f$ are incident and outgoing angles shown in Fig~\ref{fig:samplesetup}, and $\bq_{\parallel}=(q_x,q_y)$. The DWBA form factor $G(\mathbf{q}_{\parallel},\alpha^i,\alpha^f)$ is
\begin{equation}
\label{eqn:dwba}
G(\mathbf{q}_{\parallel},\alpha^i,\alpha^f)=\sum_{m=1}^{4}D^mF(q_z^m,\mathbf{q}_{\parallel})
\end{equation}

\noindent where the $D^m$ are defined as:

\begin{align}
\nonumber D^1&=T^fT^i,\indent D^2=R^fT^i, \\
D^3&=T^fR^i,\indent D^4=R^fR^i
\label{eqn:fresnel}
\end{align}

\noindent where $T^i,T^f,R^i,R^f$ are Fresnel transmission and reflection coefficients for reflectivity at the interface between the top layer and substrate. Details about the coefficients can be found in Ref.~\cite{tolan1999x}. $\mathbf{q}^m$ has been split into components $q_z^m$ and $\mathbf{q}_{\parallel}$,

\begin{align}
\nonumber \mathbf{q}^1=(\mathbf{q}_{\parallel},q_z^1), \indent \mathbf{q}^2&=(\mathbf{q}_{\parallel},q_z^2), \\
\mathbf{q}^3=(\mathbf{q}_{\parallel},q_z^3), \indent \mathbf{q}^4&=(\mathbf{q}_{\parallel},q_z^4)
\end{align}

\noindent and

\begin{align}
\nonumber q_z^1&=k_z^f-k_z^i, &&q_z^2=-k_z^f-k_z^i, \\
q_z^3&=-q_z^2, &&q_z^4=-q_z^1
\label{eqn:qz}
\end{align}

\noindent Note that, the value of $k_z^i$ is negative, since the vector $\mathbf{k}^i$ is pointing down.

The complete wavevector transfer~\cite{tolan1999x,venkatakrishnan2016multi} from the angles shown in Fig.~\ref{fig:samplesetup} is defined by the 6 elements of the left-hand-side vector:

\begin{equation}
\begin{bmatrix}
q_x \\
q_y \\
k_{z}^i \\
k_{z}^f \\
k_{z,sub}^i \\
k_{z,sub}^f
\end{bmatrix}
=k_0
\begin{bmatrix}
\cos(\alpha^f)\cos(\chi)-\cos(\alpha_i) \\
\sin(\chi)\cos(\alpha^f) \\
-\sqrt{n_{top}^2-\cos^2(\alpha^i)} \\
\sqrt{n_{top}^2-\cos^2(\alpha^f)} \\
-\sqrt{n_{sub}^2-\cos^2(\alpha^i)} \\
\sqrt{n_{sub}^2-\cos^2(\alpha^f)} \\
\end{bmatrix}
\label{eqn:wavevector}
\end{equation}

\noindent where $k_0$ is the incident X-ray wavevector, $n_{top}$ and $n_{sub}$ are the complex refractive indices of the top layer and substrate. In this paper, the top layer is air and $n_{top}=1$. $k_{z,sub}^i$ and $k_{z,sub}^f$ are the z component of wavevectors in the substrate, which together with $k_{z}^i$ and $k_{z}^f$ are used to calculate the Fresnel coefficients in Eq.~\eqref{eqn:fresnel}.

Compared to the BA form factor $F(\mathbf{q})$, the DWBA form factor $G(\mathbf{q}_{\parallel},\alpha^i,\alpha^f)$ has 4 Fourier transform terms, which correspond to 4 scattering processes. It is interesting to note that the DWBA automatically involves negative values of $q_z$ in the Fourier transforms of the electron density of the object. These are generally not directly measured in a conventional GISAXS experiment. Details of the 4 scattering processes are illustrated in Fig.(41) in Ref.~\cite{renaud2009probing}.

Note that, if the sample is a uniform thin film (2D sample), DWBA is not needed. 
Because in this case, the Fourier transform of the electron density is independent of $q_z$, $F(q_x,q_y,q_z^1)=F(q_x,q_y,q_z^2)=F(q_x,q_y,-q_z^2)=F(q_x,q_y,-q_z^1)$, and thus the DWBA form factor is proportional to the BA form factor.
If the sample has a uniform finite thickness~\cite{sun2012three}, $(q_x,q_y)$ and $q_z$ are separable, $F(q_x,q_y,q_z)=F_1(q_x,q_y)F_2(q_z)$. The BA form factor in the $(q_x,q_y)$ plane is independent of that in the $q_z$ direction.

\section{Coherent Diffraction Imaging Algorithm}
\begin{figure*}[htbp]
\centering

\begin{subfigure}{1\textwidth}
\caption{}
\includegraphics[scale=.4]{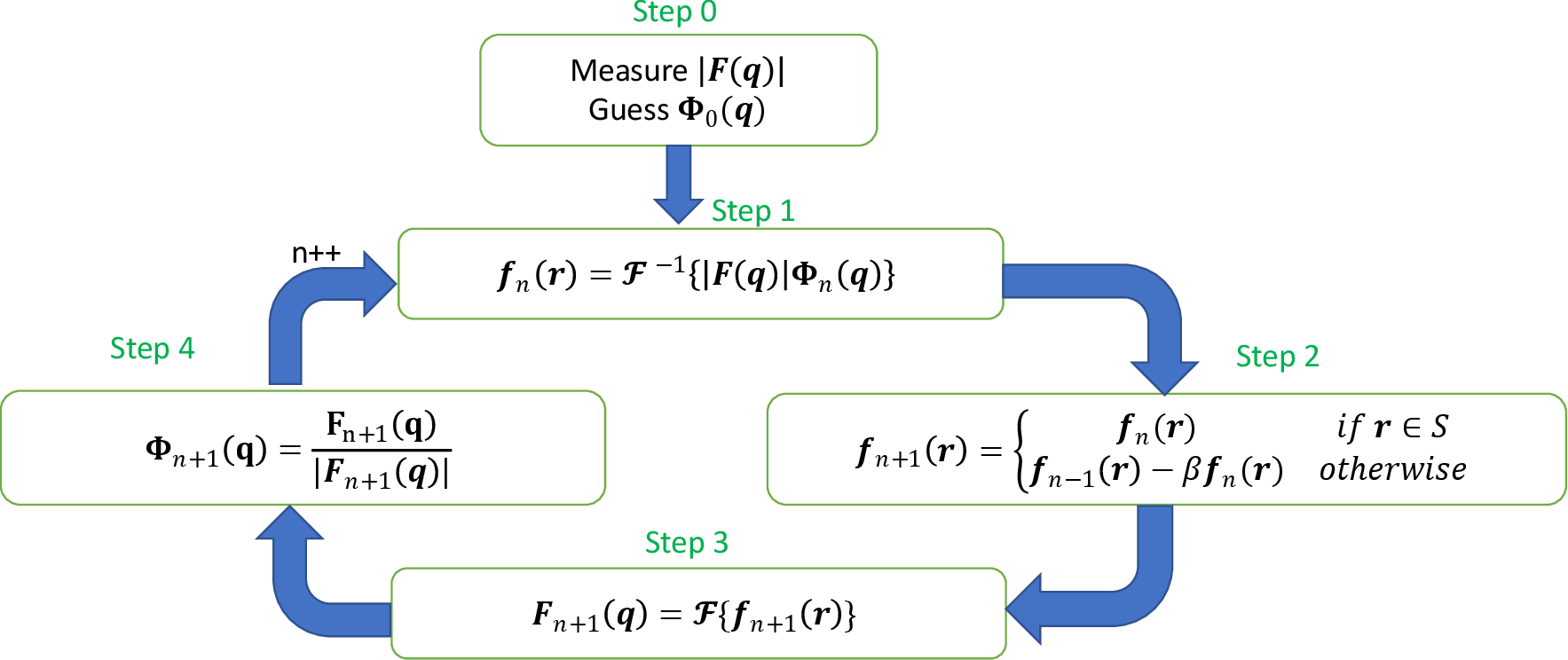}
\label{fig:baflow}
\end{subfigure}

\begin{subfigure}{1\textwidth}
\caption{}
\includegraphics[scale=.4]{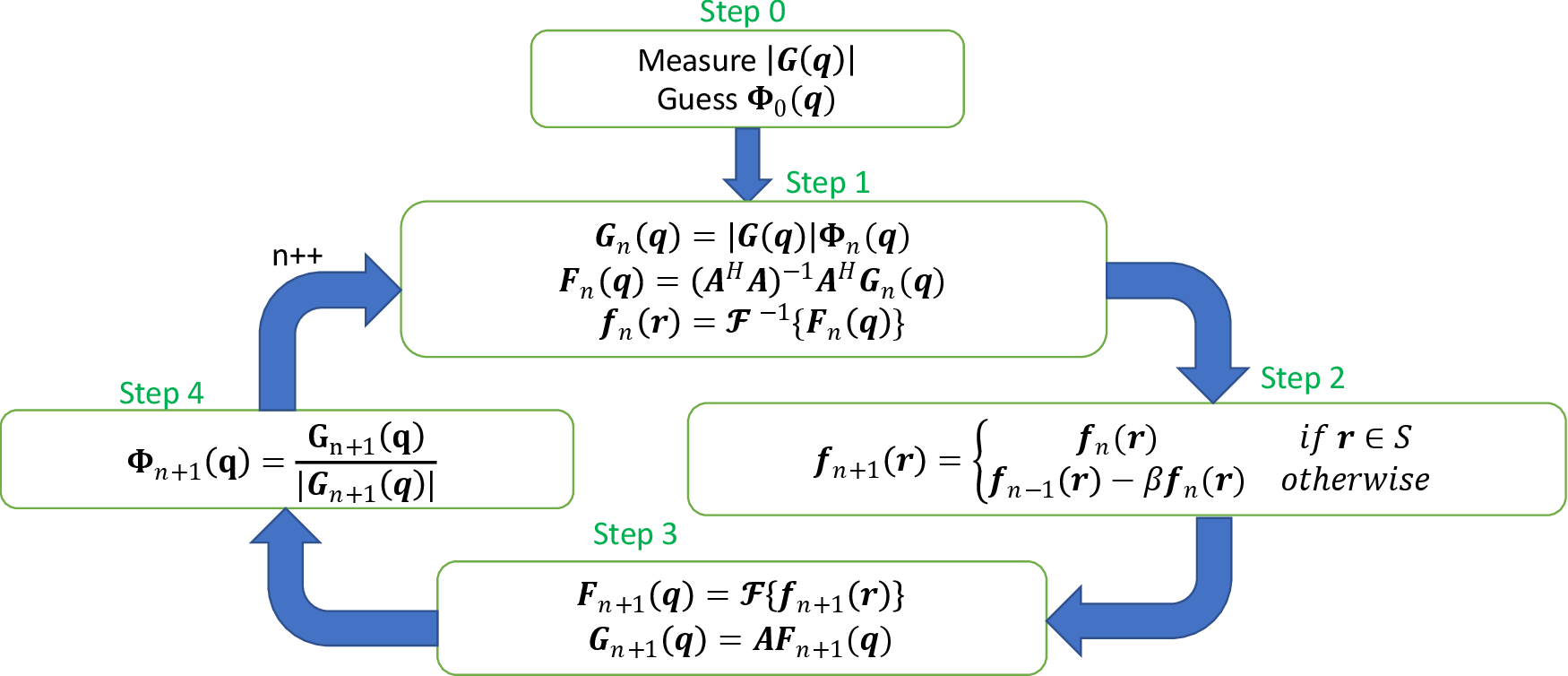}
\label{fig:dwbaflow}
\end{subfigure}


\caption{\subref{fig:baflow} regular CDI iterative steps~\cite{welch2015coherent}. \subref{fig:dwbaflow} DWBA-CDI iterative steps proposed in this paper. For both figures, S in step 2 is the set of elements inside the support, and n is the iteration number. A value of the feedback parameter of
$\protect\beta$=0.9 is used. $\protect |F(\mathbf{q})|$
in \subref{fig:baflow} and
$\protect |G(\mathbf{q})|$ in \subref{fig:dwbaflow} in step 1 are the measured scattering magnitudes in step 0. $\protect |G(\bq)|$ is short of $\protect |G(q_x,q_y,\alpha^i,\alpha^f)|$.
}
\label{fig:flowdiagram}
\end{figure*}

Several iterative algorithms have been developed to retrieve the phases of the scattering magnitudes $|F(\bq)|$ based on the principle of oversampling of the scattered intensities of the diffraction patterns~\cite{marchesini2007invited}. Fig.~\ref{fig:baflow} is an example of the Hybrid-Input-Output (HIO)~\cite{fienup1982phase} algorithm. By iterating step 1 to 4 in Fig.~\ref{fig:baflow}, the phases $\Phi(\bq)$ will converge in a few hundred iterations. Other algorithms, such as Error Reduction (ER)~\cite{gerchberg1972practical,fienup1982phase,levi1984image}, Relaxed Averaged Alternating Reflection~\cite{luke2004relaxed} and Difference Map~\cite{elser2003phase} replace the corresponding equations in step 2~\cite{marchesini2007invited}. In this paper, we combined the HIO with the ER. Details are given in the Simulation section.

The algorithm shown in Fig.~\ref{fig:baflow} first needs a random guess of the phases of the scattering magnitudes in step 0; the electron density $f(\mathbf{r})$ is calculated in step 1,  and then forced to satisfy the constraints in object space in step 2; new Fourier transform is obtained in step 3; the phases of the new Fourier transform are kept while the magnitudes are discarded in step 4. After many iterations from step 1 to 4, the phases of the BA form factor $F(\mathbf{q})$ are retrieved; however, it fails near grazing incidence because of the presence of the reflected wave from the substrate as already discussed in the introduction and DWBA section. Instead, we can use a similar algorithm to retrieve the phase of the DWBA form factor, except for the complication that to relate it to the Fourier transform of the object density, we need to solve for the 4 different terms in Eq.~\eqref{eqn:dwba}.

We proposed a matrix method to solve the more complicated DWBA form factor, which has 4 Fourier transform terms with different $q_z$'s. The matrix method is based on the fact that $q_z^1$ to $q_z^4$ are partially overlapped when we have multiple incident and outgoing angles; for example, $q_z^1(\alpha^i=0.1deg,\alpha^f=0.2deg)=q_z^2(\alpha^i=0.4deg,\alpha^f=0.1deg)$. Given enough incident and outgoing angle pairs, we can relate the DWBA form factor to the BA form factor (Fourier transform of the electron density) via a matrix, and use the matrix to solve the BA form factor in terms of the DWBA form factor (the form factor is a data array, not a scalar). The iterative algorithm is modified to be in Fig.~\ref{fig:dwbaflow}. Steps 2 and 4 remain the same while steps 1 and 3 are replaced with the combination of Fourier transform and DWBA matrix multiplications.

To solve the matrix $\BA$ shown in Fig.~\ref{fig:dwbaflow}, which is crucial to this DWBA-CDI algorithm, let us expand Eq.~\eqref{eqn:dwba}:

\begin{align}
\label{eqn:dwbap}
\nonumber G&=D^1(k_z^i,k_z^f)F(q_x,q_y,q_z^1) \\
\nonumber & +D^2(k_z^i,k_z^f)F(q_x,q_y,q_z^2) \\
\nonumber & +D^3(k_z^i,k_z^f)F(q_x,q_y,-q_z^2) \\
& +D^4(k_z^i,k_z^f)F(q_x,q_y,-q_z^1)
\end{align}

We assume we have the scattering magnitudes (the DWBA form factor magnitudes at GISAXS geometry) $|G(q_x,q_y,k_z^i,k_z^f)|$ at equally spaced $q_x,q_y,q_z^1$ to $q_z^4$, which can be achieved through interpolation in the $q_x,q_y$ directions, by rotating the sample azimuthally. We have to choose suitable incident and outgoing angles to equally space $q_z^1$ to $q_z^4$. To be clear, $G(q_x,q_y,\alpha^i,\alpha^f)$ is also written as $G(q_x,q_y,k_z^i,k_z^f)$ in the context; so is $D^m(k_z^i,k_z^f)$ and $D^m(\alpha^i,\alpha^f)$.

Given finite number of $(\alpha^i,\alpha^f)$ pairs, $\pm q_z^1$ and $\pm q_z^2$ lie in the range of $[-\frac{N_z}{2},\frac{N_z}{2}-1]$, where $N_z$ is the total number of discrete $q_z$'s. To be clear, $\pm q_z^1$ and $\pm q_z^2$ have a unit of $\Delta q_z$, which is defined in the supplementary materials. $\BF(q_x,q_y,\pm q_z^1)$ and $\BF(q_x,q_y,\pm q_z^2)$ are subsets of $\BF_{N_z\cdot 1}$, defined in Eq.~\eqref{eqn:ffvector}. $\BG$ and $\BF$ are defined as column vectors:

\begin{align}
\nonumber \BG&=
\begin{bmatrix}
G(q_x,q_y,\alpha_{1}^i,\alpha_{1}^f) \\
\vdots \\
G(q_x,q_y,\alpha_{J}^i,\alpha_{J}^f)
\end{bmatrix}_{J\cdot 1}
\\
\BF&=
\begin{bmatrix}
F(q_x,q_y,-\frac{N_z}{2}) \\
\vdots \\
F(q_x,q_y,\frac{N_z}{2}-1)
\end{bmatrix}_{N_z\cdot 1}
\label{eqn:ffvector}
\end{align}

\noindent where $J$ is the total number of $(\alpha^i,\alpha^f)$ pairs (incident and outgoing angle pairs in Fig.~\ref{fig:samplesetup}), and $N_z$ is the total number of possible $q_z$'s. We have the matrix form of Eq.~\eqref{eqn:dwbap} for each $(q_x,q_y)$ pair:

\begin{equation}
\label{eqn:dwbamatrix}
\BG_{J\cdot 1}=\BA_{J\cdot N_z}\BF_{N_z\cdot 1}
\end{equation}

$\BA_{J\cdot N_z}$ can be calculated by inserting Eq.~\eqref{eqn:dwbap} into Eq.~\eqref{eqn:dwbamatrix}. $\BA_{J\cdot N_z}$ is independent of $(q_x,q_y)$, while $\BG$ and $\BF$ are not. That is, we need to calculate the matrix \textbf{A} only once for all the $(q_x,q_y)$ involving those particular values of $q_z$. Given the magnitudes and phases of $\BG$, Eq.~\eqref{eqn:dwbamatrix} can then be inverted via the least square method~\cite{wiki:Least_squares} if $J\geq N_z$ and matrix $\BA$ is non-singular:

\begin{equation}
\label{eqn:lsm}
\BF=(\BA^H \BA)^{-1}\BA^H\BG
\end{equation}
\noindent where $\BA^H$ is the conjugate transpose (Hermitian transpose) of matrix $\BA$.


However, the least square method in general does not work in this case. Because at incident and outgoing angles greater than twice of the substrate's critical angle (maximum of incident and outgoing angles have to be large enough to satisfy the condition $J>N_z$), 
$D^4$ becomes very small and matrix $\BA$ is singular. Therefore, $|\BA^H \BA|\approx0$ and $(\BA^H \BA)^{-1}$ is ill defined.

We fixed the problem by introducing Friedel's Law, which doubles the number of equations without changing the number of unknown variables. In most cases, the imaginary part of the sample electron density is much smaller than the real part, and may be neglected. In such cases, Friedel's Law states:

\begin{equation}
\label{eqn:friedel}
F(-q_x,-q_y,-q_z)=F^*(q_x,q_y,q_z)
\end{equation}

\noindent where $F^*$ is the complex conjugate of $F$. By the way, Friedel's Law is also necessary for 3D diffraction imaging in reflection geometry even using the BA. Because it gives the scattering information in the negative $q_z$ direction, which BA does not.

Here, let us use the new notation at fixed $q_x,q_y$:

\begin{align}
\nonumber G^+&=G(q_x,q_y,\alpha^i,\alpha^f), \\
G^-&=G(-q_x,-q_y,\alpha^i,\alpha^f)
\end{align}

\noindent where $G^+$ is the same as the $G$ in Eq.~\eqref{eqn:dwbap} and

\begin{align}
\label{eqn:dwbam}
\nonumber G^-= & D^1(k_z^i,k_z^f)F^*(q_x,q_y,-q_z^1) \\
\nonumber  + & D^2(k_z^i,k_z^f)F^*(q_x,q_y,-q_z^2) \\
\nonumber  + & D^3(k_z^i,k_z^f)F^*(q_x,q_y,q_z^2) \\
 + & D^4(k_z^i,k_z^f)F^*(q_x,q_y,q_z^1)
\end{align}

According to Eq.~\eqref{eqn:dwbap} and Eq.~\eqref{eqn:dwbam}, both $\BG^+$ and $\BG^-$ are functions of $Re[\BF_{N_z\cdot 1}]$ and $Im[\BF_{N_z\cdot 1}]$. By calculating the real and imaginary parts of $\BG^+$ and $\BG^-$, we get:

\begin{equation}
\label{eqn:dwbacdif}
\begin{bmatrix}
Re[\BG^+] \\
Im[\BG^+] \\
Re[\BG^-] \\
Im[\BG^-]
\end{bmatrix}
=
\begin{bmatrix}
\BA_1 & \BA_2 \\
\BA_3 & \BA_4 \\
\BA_5 & \BA_6 \\
\BA_7 & \BA_8
\end{bmatrix}
\begin{bmatrix}
Re[\BF] \\
Im[\BF]
\end{bmatrix}
\end{equation}

\noindent Define:

\begin{equation}
\label{eqn:dwbamtx}
\BA_r=
\begin{bmatrix}
\BA_1 & \BA_2 \\
\BA_3 & \BA_4 \\
\BA_5 & \BA_6 \\
\BA_7 & \BA_8
\end{bmatrix}_{4J\cdot 2N_z}
\end{equation}

The matrices $\BA_1$ to $\BA_8$ are real and have size of $J\cdot N_z$. $\BA_r$ is not singular, if $J$ is big enough. While the matrix $\BA$ in Eq.~\eqref{eqn:dwbamatrix} is complex. Details of calculating $\BA_1$ to $\BA_8$ are given in supplementary document. In Fig.~\ref{fig:dwbaflow}, we actually used Eq.~\eqref{eqn:dwbacdif} instead of Eq.~\eqref{eqn:dwbamatrix}.

Finally, given $|G(\mathbf{q})|$ and retrieving the corresponding phases $\Phi(\mathbf{q})$ via DWBA-CDI algorithm, $F(\mathbf{q})$ can be calculated via the least square method, so is the electron density $f(\mathbf{r})$ (the image of the object).

\section{Simulation Result}
\begin{figure}[ht]
\centering

\begin{subfigure}[b]{.475\columnwidth}
\caption{}
\label{fig:o3d}
\includegraphics[scale=.3]{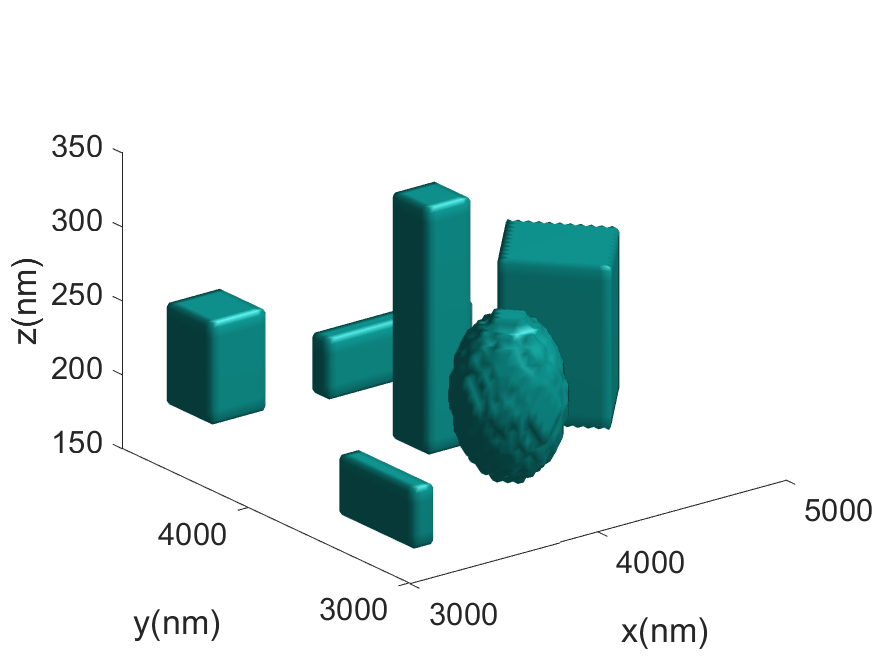}
\end{subfigure}
\hfill
\begin{subfigure}[b]{.475\columnwidth}
\caption{}
\label{fig:r3ddwba}
\includegraphics[scale=.3]{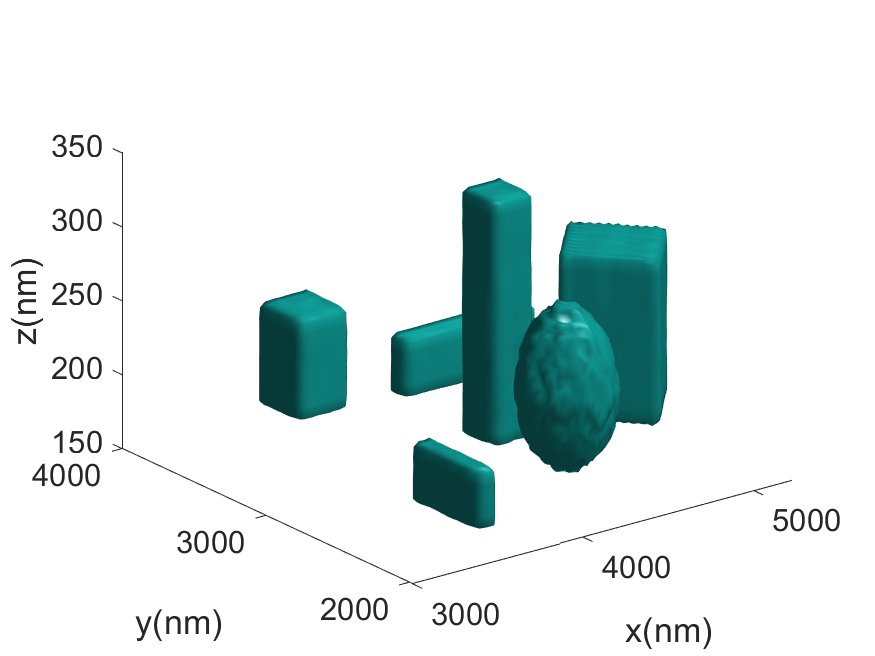}
\end{subfigure}

\begin{subfigure}[b]{.475\columnwidth}
\caption{}
\label{fig:r3dba}
\includegraphics[scale=.3]{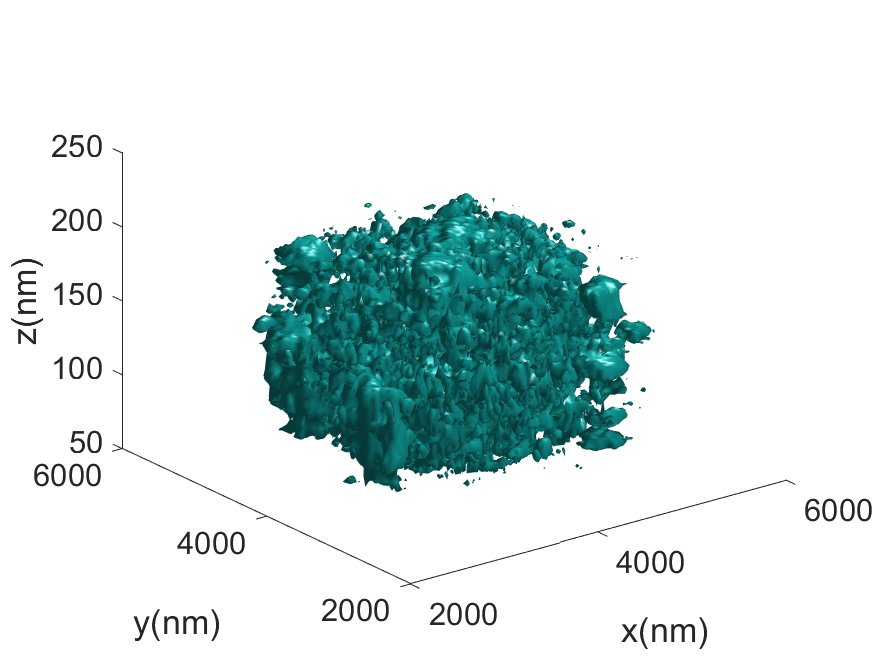}
\end{subfigure}
\hfill
\begin{subfigure}[b]{.475\columnwidth}
\caption{}
\includegraphics[scale=.2]{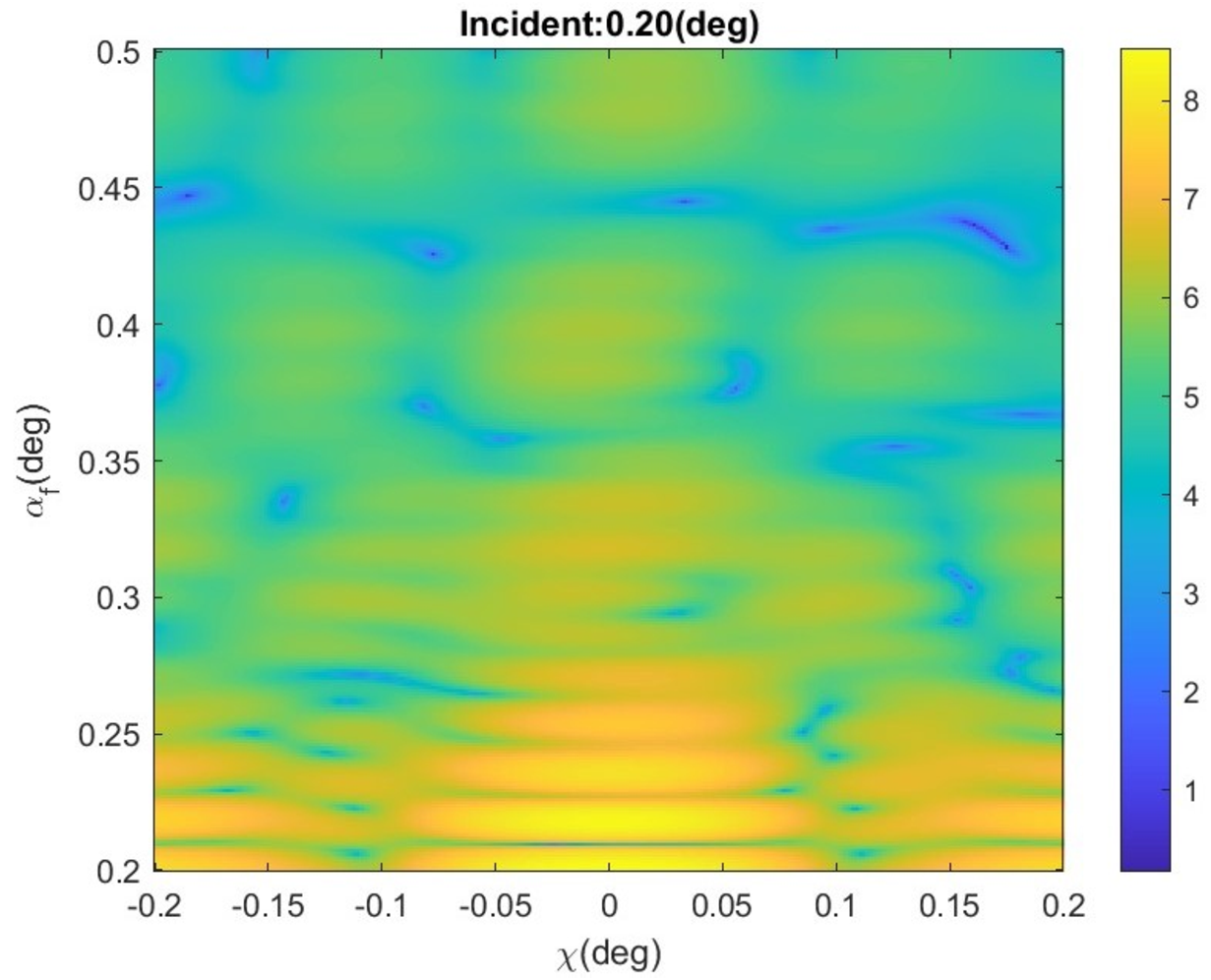}
\label{fig:diffpattern}
\end{subfigure}

\caption{
3D angle-view of the \subref{fig:o3d} original sample (substrate not shown here), \subref{fig:r3ddwba} reconstructed sample using the DWBA and \subref{fig:r3dba} reconstructed sample using the BA;
\subref{fig:diffpattern} A diffraction pattern example of the sample at the incident angle 0.2 degree.
}

\label{fig:cdiresult}
\end{figure}

We chose a sample made up of random solid shapes made of gold, sitting on a Silicon substrate in a certrain area, shown in Fig.~\ref{fig:samplesetup}. Fig.~\ref{fig:o3d} represents the 3D angle-view of the sample. For simplicity, we set the electron density of Si to be $1$ in the code, and $6.5$ for gold for the reconstruction purpose. However, for the DWBA calculation, the actual electron density and refractive index of Silicon is used to calculate the reflection and transmission coefficients in Eq.~\eqref{eqn:fresnel} and ~\eqref{eqn:wavevector}. The X-ray energy was chosen as $8.04keV$, with a wavelength of $1.54\ANG$.

For the simulation, we have assumed that the sample is small enough to always remain inside the footprint of the beam and that the photon flux is uniform across the beam. 
The reconstruction result assuming Gaussian beam is included in the Supplementary materials.
The diffuse scattering from the Si surface was taken to be negligible. And there is a beamstop (not shown in the figure) on the detector center, to block the specular scattering from Si surface. The beamstop size in reciprocal space is $\Delta q_x=\Delta q_y\approx 2\cdot 10^{-4}\ANG^{-1},\Delta q_z\approx10^{-3}\ANG^{-1}$ (1 pixel size in the $q_x,q_y,q_z$ directions). The error contributed from the beamstop is very small and neglected here. The contribution of error from the beamstop is well discussed in Ref.~\cite{miao2005quantitative}.

Incident angles were taken to be from $0.01^{\circ}$ to $0.63^{\circ}$ with a step of $0.02^{\circ}$. The Si critical angle is $0.223^{\circ}$. For each incident angle, the sample is rotated azimuthally in 180 steps from $0^{\circ}$ to $180^{\circ}$. After getting 32*180 diffraction patterns, we interpolate them to a 3D diffraction pattern in reciprocal space. Here we used 180 rotations per incident angle to guarantee the quality of interpolation in the reciprocal space. Less rotations may result in bad quality of image reconstruction.

To speed up the program in Fig.~\ref{fig:dwbaflow}, we applied the non-negative constraint (ER) of electron density in step 2 every 60 iterations~\cite{marchesini2007invited} and update the support every 100 iterations following the Shrink-wrap method~\cite{marchesini2003x}. Total number of iterations is 1,000. The reconstructed 3D image is shown in Fig.~\ref{fig:r3ddwba}. In comparison, we also reconstructed the image using the BA, shown in Fig.~\ref{fig:r3dba}. We also tried the Oversampling Smoothness(OSS) algorithm~\cite{rodriguez2013oversampling}, which does not perform as well (see the supplementary document for comparison between HIO-ER which is used in the paper and HIO-OSS).

We added Gaussian noise to the simulated diffraction pattern, defined as:
\begin{equation}
noise = \frac{signal}{SNR}*random
\end{equation}

\noindent where SNR is the signal-to-noise ratio, we chose $SNR=10$ in the paper; $random$ are Gaussian random numbers $\sim N(0,1)$. An example of diffraction pattern at incident $\alpha^i=0.2^{\circ}$ is shown in Fig.~\ref{fig:diffpattern}. To characterize the reconstructions, we used the error and convergence functions, which are calculated as:

\begin{align}
\nonumber Error=\sqrt{\frac{\sum_{\bq}(|F_{orig}(\bq)|-|F_r(\bq)|)^2}{\sum_{\bq}(|F_{orig}(\bq)|+|F_r(\bq)|)^2}} \\
Convergence(i) = \sqrt{\frac{\sum_{\bq}|F_r^i(\bq)-F_r^{i-1}(\bq)|^2}{\sum_{\bq}|F_r^i(\bq)+F_r^{i-1}(\bq)|^2}}
\label{eqn:error}
\end{align}

\noindent where $|F_{orig}(\bq)|$ is the magnitude of the original BA form factor; $|F_r(\bq)|$ is the magnitude of the reconstructed BA form factor in the final iteration; $F_r^i(\bq)$ is the BA form factor in the $i_{th}$ iteration. In brief, $Error$ is the difference between the original and final reconstructed images; while $Convergence(i)$ is the difference between the $i-1_{th}$ and $i_{th}$ iteration. The reason we use the magnitude of BA form factor to calculate the error is that translational shift of the whole objects between original and final reconstructed images does not affect the error. Typically, the $Convergence$ in the final iteration (which is iteration 1,000 in our case) is positively correlated with the $Error$ according to our simulation. We ran the code multiple times and chose the one with smallest final $Convergence$, and also the smallest error, which is $2\%$. More details about the $Error$ and $Convergence$ are given in the supplementary materials. 

The reconstructed shapes using the DWBA in Fig.~\ref{fig:r3ddwba} is clear with sharp edges and of the same sizes as the original. The relative distances between shapes remain the same as the original image. The relative locations of the shapes in the z direction are correct. However, reconstruction using the BA is totally a failure; and the iterations do not converge. Note that, people have successfully reconstructed images in reflection geometry before~\cite{zhu2015ptychographic,sun2012three}; but these have been quasi-2-dimensional. This is the first time that we have shown we can reconstruct detailed structure of the complex objects in the z direction using reflection geometry. Note that, the reconstruction result may have an arbitrary translational shift of the original image in the x and y directions, due to the fact that arbitrary translational shift of the whole objects does not affect the diffraction pattern in these directions. However the substrate acts as a marker in the z direction so there is no arbitrary translation of the image in this direction.

\begin{figure}
\centering

\begin{subfigure}{.475\columnwidth}
\caption{}
\label{fig:ob3d}
\includegraphics[scale=.3]{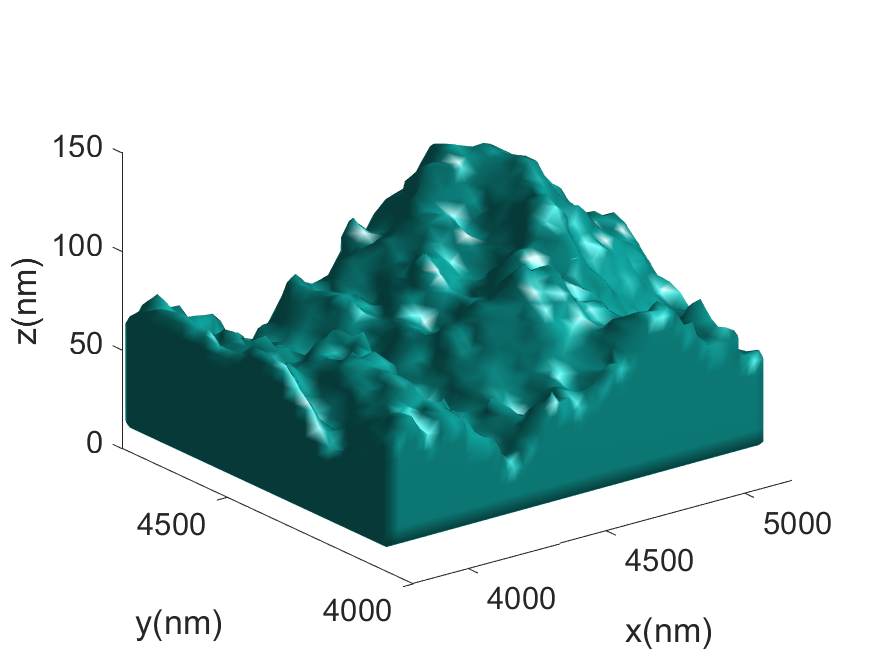}
\end{subfigure}
\hfill
\begin{subfigure}{.475\columnwidth}
\caption{}
\label{fig:rb3ddwba}
\includegraphics[scale=.3]{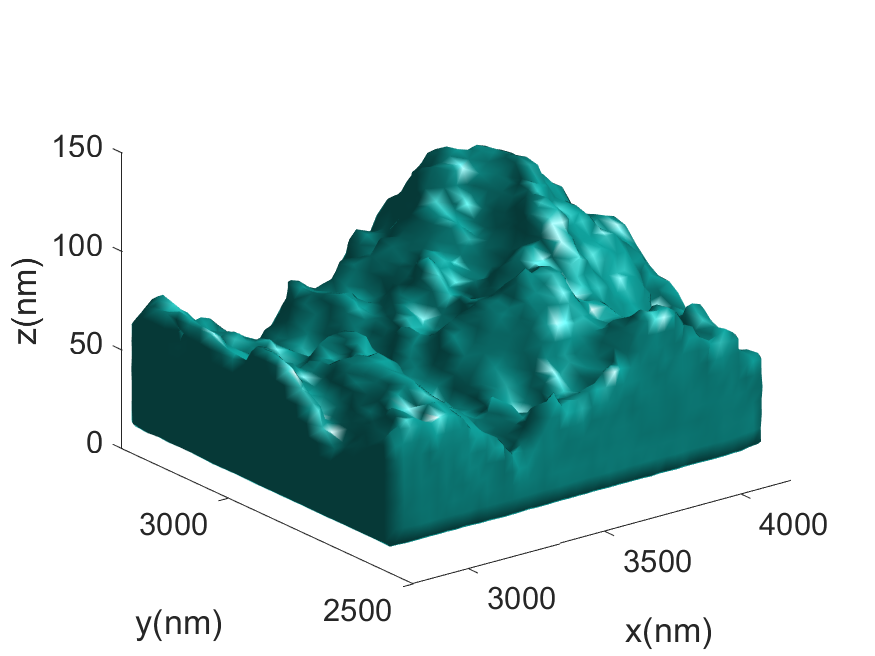}
\end{subfigure}

\caption{
\subref{fig:ob3d} 3D angle-view of the \subref{fig:ob3d} original box with rough top surface (substrate not shown here) and \subref{fig:rb3ddwba} reconstructed sample using the DWBA.
}

\label{fig:roughbox}
\end{figure}

Other than the isolated shapes, we also tested the reconstruction of a single box with a rough top surface, with an error of $9\%$, as shown in Fig.~\ref{fig:roughbox}. The algorithm works for a single rough surface with limitations, i.e the rough surface needs to be smaller than the footprint of the beam, and the roughness needs to be bigger than the resolution in the z direction in object space (which is limited by the largest incident and outgoing angles).

By the definition of discrete Fourier transform Eq.~\eqref{eqn:ft}, object space resolution (pixel size in object space), the maximum object space size, the reciprocal space resolution (pixel size in the reciprocal space), and the maximum reciprocal space size is related as (in the $z$ and $q_z$ direction as an example):

\begin{align}
\nonumber q_{z,max}\Delta z=\pi \\
z_{max}\Delta q_z=2\pi
\label{eqn:resolution}
\end{align}

\noindent where $q_z\in [-q_{z,max},q_{z,max}]$ and $z\in [0,z_{max}]$. $\Delta q_z$ is the resolution in the $q_z$ direction in the reciprocal space, $q_{z,max}$ is the maximum of $q_z$, and the same for the z direction in object space. From the above equation, it is easy to find out that the resolution in the z direction is determined by the maximum of $q_z$, which is determined by Eq.~\eqref{eqn:qz} and ~\eqref{eqn:wavevector}. In our simulation, $n_{top}=1$, $q_{z,max}=k_0(sin\alpha^i_{max}+sin\alpha^f_{max})$. In short, the resolution in the z direction is determined by the largest incident and outgoing angles. In a real experiment, the diffraction intensity at high incident and outgoing angles is too low to be detected, thus limit the z resolution to be around $10nm$, while the theoretical resolution is about the wavelength of the incident X-ray (typically smaller than $1nm$ for hard X-rays).

Scanning CDI/ptychography would be the next step in the development of this method , but is complicated by the highly anisotropic geometry of GISAXS experiments and of the footprint on the sample, and the necessity to get diffraction patterns from many in-plane orientations. We believe these problems are solvable and are working on them. 

\section{Conclusion}
The algorithm succeeds in reconstructing the 3D image at grazing incidences, while regular CDI using BA does not. As we can reach to much smaller incident and outgoing angles, as well as $|q_z^2|<|q_z^1|$, the minimum of $q_z$ in DWBA is much smaller. That is, DWBA is able to reconstruct much thicker sample. BA cannot in principle reconstruct objects thicker than around 5nm.

The results obtained from computer simulations of the scattering and reconstruction are very encouraging in showing that the method works. Verification with real experiments is planned.

%
%
%
%
%
%
%

\section*{Acknowledgement}
The authors are thankful to Zhang Jiang, Jin Wang and Jianwei Miao for valuable discussions.
This work was supported by grant No. DE-SC0003678 from the Division of Basic Energy Sciences of the Office of Science of the U.S. Dept. of Energy.

\bibliography{mybib}
\bibliographystyle{unsrt}

\end{document}


\maketitle

\section*{Derivation of the DWBA Matrix}
Here we show how to calculate $\BA_1\sim \BA_8$ from Eq.(9) and (15) in the main paper. Let us first rewrite the Eq. (9) and (14) in the main paper, at fixed $q_x,q_y$:
\begin{align}
\label{eqn:dwbap}
G^+=D^1F[q_z^1]+D^2F[q_z^2]+D^3F[-q_z^2]+D^4F[-q_z^1] \\
\label{eqn:dwbam}
G^-=D^1F^*[-q_z^1]+D^2F^*[-q_z^2]+D^3F^*[q_z^2]+D^4F^*[q_z^1]
\end{align}
\noindent where $G^+$, $G^-$, $D^m$ ($m=1,2,3,4$) are all functions of incident and outgoing angles $(\alpha^i,\alpha^f)$.

If there are $J$ pairs of $(\alpha^i,\alpha^f)$ at fixed $q_x,q_y$, there are $J$ equations for both $G^+$ and $G^-$. We can rewrite Eq.~\eqref{eqn:dwbap} and~\eqref{eqn:dwbam} in matrix forms. Define $\BG^+$, $\BG^-$ as vectors of size $J\cdot 1$, corresponding to the number of $(\alpha^i,\alpha^f)$, and $\BF$ as a vector of size $N_z\cdot 1$ (see the main paper Eq.(10) for the definition of $\BF$), corresponding to all the different $q_z$'s ($N_z$ is dependent of $J$). Let us define a matrix $\BD^+_{jn}$ where $j$ refers to a particular $(\alpha^i,\alpha^f)$ pair, and $n$ refers to a particular component of the column vector $\BF[q_z]$.

For every $j$, there is a corresponding $q_z^1$, $q_z^2$, $-q_z^2$, $-q_z^1$, as defined by Eq.(7) in the main paper. Let these correspond to the elements $k,m,m^{'},k^{'}$ respectively in the column vector $\BF[q_z]$. The corresponding matrix elements are:
\begin{align}
\label{eqn:dpmtx}
\nonumber \BD^+_{jk} = D^1(\alpha^i,\alpha^f) \\
\nonumber \BD^+_{jm} = D^2(\alpha^i,\alpha^f) \\
\nonumber \BD^+_{jm^{'}} = D^3(\alpha^i,\alpha^f) \\
\BD^+_{jk^{'}} = D^4(\alpha^i,\alpha^f)
\end{align}
\noindent from Eq.~\eqref{eqn:dwbap}. Where $D^1$ to $D^4$ are defined by Eq.(5) in the main paper.  All other $\BD^+_{jn}=0$ if $n\neq k,m,m^{'},k^{'}$.

Similarly, we define a matrix $D^-_{jn}$ with
\begin{align}
\label{eqn:dmmtx}
\nonumber \BD^-_{jk} = D^4(\alpha^i,\alpha^f) \\
\nonumber \BD^-_{jm} = D^3(\alpha^i,\alpha^f) \\
\nonumber \BD^-_{jm^{'}} = D^2(\alpha^i,\alpha^f) \\
\BD^-_{jk^{'}} = D^1(\alpha^i,\alpha^f)
\end{align}
\noindent All other $\BD^-_{jn}=0$ if $n\neq k,m,m^{'},k^{'}$. Then we can write Eq.~\eqref{eqn:dwbap} and \eqref{eqn:dwbam} as
\begin{align}
\nonumber \BG^+=\BD^+\BF \\
\BG^-=\BD^-\BF^*
\end{align}
\noindent where $\BF^*$ is the complex conjugate of $\BF$. Both $\BG^+$ and $\BG^-$ are functions of $Re(\BF)$ and $Im(\BF)$. Next, let us get the real and imaginary part of $\BG^+$ and $\BG^-$. We assume $|k_z^i|\geq |k_z^f|$ (same as $q_z^2\geq 0$) in the context.

\begin{align}
\nonumber Re(\BG^+)&=Re(\BD^+)Re(\BF)-Im(\BD^+)Im(\BF) \\
\nonumber  Im(\BG^+)&=Im(\BD^+)Re(\BF)+Re(\BD^+)Im(\BF) \\
\nonumber Re(\BG^-)&=Re(\BD^-)Re(\BF)+Im(\BD^-)Im(\BF) \\
\nonumber  Im(\BG^-)&=Im(\BD^-)Re(\BF)-Re(\BD^-)Im(\BF)
\end{align}
Define:
\begin{align}
\BA_1&=Re(\BD^+), \quad \BA_2=-Im(\BD^+) \\
\BA_3&=Im(\BD^+), \quad \BA_4=Re(\BD^+) \\
\BA_5&=Re(\BD^-), \quad \BA_6=Im(\BD^-) \\
\BA_7&=Im(\BD^-), \quad \BA_8=-Re(\BD^-)
\end{align}

Finally, we get Eq.'s (16) and (17) in the main paper. To be clear, the real and imaginary part of $\BG^+$ and $\BG^-$ have dimensions of $J\cdot 1$. While the real and imaginary part of $\BF$ have dimensions of $N_z\cdot 1$. $\BA_1$ to $\BA_8$ have dimensions of $J\cdot N_z$.

Below is an example of calculating the matrix $\BD_{mat}^1$. To simplify the example, we choose 4 incident and outgoing angles (see Table.~\ref{tab:dwbaangles}). We assume $\alpha^i\geq \alpha^f$ ($q_z^2\geq 0$); because swapping incident and outgoing angles does not change the DWBA form factor. According to the table, there are 10 valid angle pairs, which means $J=10$. There are 17 different $q_z$'s, ranging from -8 to 8, which means $N_z=17$.

\vspace{0.25in}
\begin{table}[!ht]
\caption[Angle pairs and corresponding wavetransfers]{ Angle pairs and corresponding wavetransfers.
The incident and outgoing angles, with a unit of $\Delta\alpha$. The wavetransfer $q_z$'s have a unit of $\Delta q_z=k_0\sin\Delta\alpha$, where $k_0$ is the incident X-ray beam wavevector.
}

\begin{center}
\begin{tabular}[c]{cccccc}

\hline
$\alpha^i$($\Delta\alpha$) & $\alpha^f$($\Delta\alpha$) & $q_z^1$($\Delta q_z$) & $q_z^2$($\Delta q_z$) & $q_z^3$($\Delta q_z$) & $q_z^4$($\Delta q_z$) \\

\hline
1 & 1 & 2 & 0 & 0 & -2 \\

  & 2 & -- & -- & -- & -- \\

  & 3 & -- & -- & -- & -- \\

  & 4 & -- & -- & -- & -- \\

2 & 1 & 3 & 1 & -1 & -3 \\

  & 2 & 4 & 0 & 0 & -4 \\

  & 3 & -- & -- & -- & -- \\

  & 4 & -- & -- & -- & -- \\
  
3 & 1 & 4 & 2 & -2 & -4 \\

  & 2 & 5 & 1 & -1 & -5 \\

  & 3 & 6 & 0 & 0 & -6 \\

  & 4 & -- & -- & -- & -- \\
  
4 & 1 & 5 & 3 & -3 & -5 \\

  & 2 & 6 & 2 & -2 & -6 \\

  & 3 & 7 & 1 & -1 & -7 \\

  & 4 & 8 & 0 & 0 & -8 \\

\hline
\end{tabular}
\end{center}
\label{tab:dwbaangles}
\end{table}

According to the first row of the table ($j=1$), we have the first element of vector $\BG^+$: $\BG^+\{1\} = D^1(1,1)\BF[2]+D^2(1,1)\BF[0]+D^3(1,1)\BF[0]+D^4(1,1)\BF[-2]$. That is, the first row of matrix $\BD^+$ has 3 non-zeros elements: $D^1(1,1)$ at the $11th$ column, corresponding to the $F[2]$ in Eq. (10) in the main paper, $D^2(1,1)+D^3(1,1)$ at the $9th$ ($F[0]$) column, and $D^4(1,1)$ at the $7th$ ($F[-2]$) column. Similarly, we can get all 10 rows of matrix $\BD^+$ and $\BD^-$.

Actually, in this simple example, we cannot use this DWBA matrix $\BA$ to get the BA form factor; because $J<N_z$ (less equations than unknown variables). In the main paper, we chose $J=272,N_z=64$, to guarantee $J>N_z$ and $|A^HA|\neq 0$. In addition, we use fast Fourier transform in the paper, which requires the $q_z$'s to be non-negative. We can assume the Fourier transform of the sample to be periodic in the $q_x,q_y$ and $q_z$ directions.

\section*{Performance of HIO-OSS and HIO-ER Algorithms}
As mentioned in the main paper, the HIO-OSS does not perferm as well as HIO-ER, which has a smaller minimum $Error$ among 100 independent runs; the latter is used in the main paper. We ran both the HIO-OSS and HIO-ER algorithms 100 times (each time we ran 1,000 iterations.) using the same diffraction patterns with $SNR=10$. Table.~\ref{tab:oss1} shows that even though HIO-OSS has a smaller average $Error$ among 100 runs; The minimum $Error$ is much greater than that of HIO-ER. The authors believe that the HIO-OSS performs better with smaller $SNR$ (bigger noise) and different type of noise with non-zero mean values, rather than Gaussian noise. Because the interpolation process averages out the Gaussian noise to some degree.

\begin{figure}[htp]
\centering
\includegraphics[scale=.9]{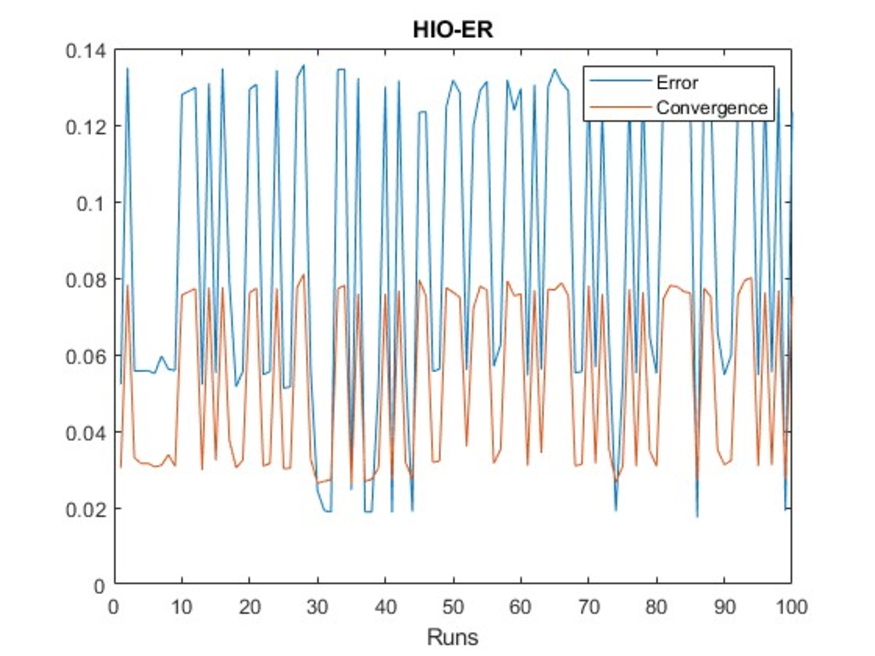}
\caption{
The relation between $\protect Error$ and $\protect Convergence$ in 100 independent runs, using HIO-ER algorithm.
}
\label{fig:rfnoise}
\end{figure}

As the original electron density (which is assumed unknown during reconstruction) is needed to calculate the $Error$ of reconstruction, we used the metric $Convergence$ to find the best reconstruction image among different runs. The $Error$ and $Convergence$ are defined in the paper. The Fig.~\ref{fig:rfnoise} shows that $Convergence$ and $Error$ are postively correlated.

\vspace{0.25in}
\begin{table}[!ht]
\caption{
Comparison between the HIO-OSS and HIO-ER algorithms. The results are from 100 independent runs at noise level $\protect SNR=10$.
}

\begin{center}
\begin{tabular}[c]{ccc}

\hline
 & $Min. Error(\%)$ & $Avg. Error(\%)$ \\
\hline
\textbf{HIO-ER} & \textbf{1.74} & \textbf{8.83} \\
HIO-OSS & 5.12 & 8.55 \\
\hline
\end{tabular}
\end{center}
\label{tab:oss1}
\end{table}

\section*{Non-uniform Incident Beam}
\begin{figure}[htp]
\centering
\includegraphics[scale=.5]{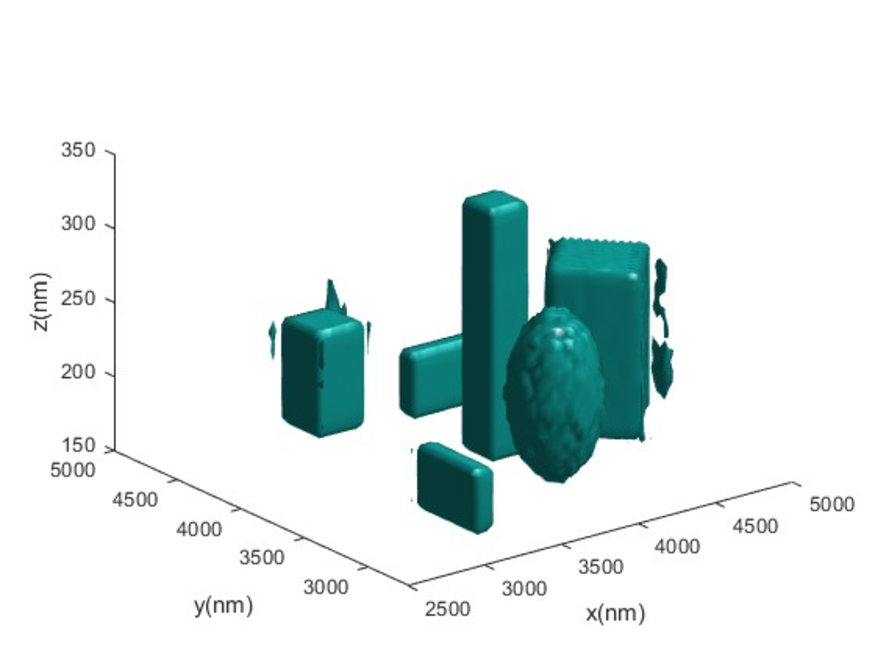}
\caption{
Reconstructed sample using DWBA with SNR=10 and non-uniform Gaussian incident beam.
}
\label{fig:gauss}
\end{figure}

The implementation also works if the incident beam is not uniform. One can also assume the incident beam has a Gaussian profile (non-uniform) in the y direction, while uniform in the x and z direction, given the sample is small in size in the x and z direction compared to the footprint of the beam. Non-uniform incident beam causes the reconstruction error from $1.74\%$ to around $10\%$, if the edge of the beam has $80\%$ intensity of its center. In Fig.~\ref{fig:gauss}, there are defects in the reconstruction image.

\section*{Implementation Details}
This section gives some details about the computer simulation.
\subsection*{Sample and Support}
The specimen in Fig.3(a) in the main paper consists of:

\begin{itemize}
    \item 1 sphere of radius 8 voxels;
    \item 5 rectangles with different length, width and height, comparable to the size of the sphere.
\end{itemize}

They are all floating/sitting on the same z plane above the bottom of the object space, in order not to break the periodic boundary condition in object space, which is necessary for discrete Fourier Transform. The whole object space has 256*256*64 voxels. The electron density of all the shapes are set to be 1, while vacuum is 0.

The shape of the initial support is ellipsoid. The major axis is 64 voxels in the x and y direction, and the minor axis is 16 voxels in the z direction. The support is loose, bigger than enough to cover the specimen. According to our study, the shape of initial support does affect the reconstruction error. Ellipsoid is better than rectangle.

\subsection*{DWBA Scattering}
The specimen in the previous subsection is in the same layer (the medium is vacuum). Diffraction patterns were collected from multiple incident angles ranging from 0.01 degree to 0.63 degree with a step of 0.02 degree (32 different incident angles). At each incident angle, the specimen is rotated along the z axis from 0 degree to 180 degree with a step of 1 degree and 180 diffraction patterns were collected. On the diffraction pattern, only the data at outgoing angles from 0.01 degree to 0.63 degree with a step of 0.02 degree were kept. There are 2 more constraints for choosing the incident and outgoing angles:
\begin{itemize}
    \item Incident angle equal to or bigger than outgoing angle
    \item The sum of incident and outgoing angles smaller than or equal to 0.65 degree
\end{itemize}
After applying those constraints, there are 272 different incident, outgoing angle pairs, corresponding to 64 different $q_z$'s. The $q_z$'s are ranging from $k_0(sin(0.01^{\circ})-sin(0.63^{\circ}))$ to $k_0(sin(0.01^{\circ})+sin(0.63^{\circ}))$ with a step of $k_0sin(0.02^{\circ})$.

The details of calculating the DWBA matrix is given in previous sections and omitted here. Gaussian noise (SNR=10) is added in this part, which has very small impact on reconstruction error. All the diffraction patterns are interpolated together to a 3D diffraction pattern. One can also assume the incident beam has a Gaussian profile (non-uniform) in the y direction, while uniform in the x and z direction. Non-uniform incident beam causes the reconstruction error from $1.74\%$ to around $10\%$, if the edge of the beam has $80\%$ intensity of its center.

This part is suggested  running on GPU.

\subsection*{Image Reconstruction}
The center voxel of the 3D diffraction pattern from previous section is blocked. The value of that voxel is updated in every iteration. We followed the Hybrid-Input-Output method. In every 60 iterations, the electron density of the specimen is forced to be real by taking the absolute value. In every 100 iterations, the support is updated following the shrink-wrap method; the threshold is set to be 0.13 while the electron density of the specimen is 1. Typically, 1000 iterations is enough.